\documentclass[aps,prl,twocolumn,superscriptaddress]{revtex4}

\usepackage{graphicx}
\usepackage{dcolumn}
\usepackage{bm}
\usepackage{amsmath}
\usepackage{amssymb}
\usepackage{latexsym}
\usepackage{epsfig}
\usepackage{amsbsy}
\usepackage{array}
\usepackage{amssymb}
\usepackage{setspace}
\usepackage{bm}

\def\sint{\ifmmode{- \!\!\!\!\!\! \int}
    \else{\hbox{$- \!\!\!\! \int \ $}}\fi}

\begin{document}

\preprint{Physical Review Letters}

\title{Ultrasensitive $N$-photon interferometric autocorrelator}

\author{Zili Zhou\footnote{Author to whom correspondence should be
addressed; electronic mail:z.zhou@tue.nl.} }
\author{Giulia Frucci}
\affiliation{COBRA Research Institute, Eindhoven University of Technology, P.O. Box 513, 5600 MB Eindhoven, The Netherlands}
\author{ Francesco Mattioli}
\author{ Alessandro Gaggero}
\author{ Roberto Leoni }
\affiliation{Istituto di Fotonica e Nanotecnologie (IFN), CNR, via Cineto Romano 42, 00156 Rome, Italy}
\author{Saeedeh Jahanmirinejad}
\affiliation{COBRA Research Institute, Eindhoven University of Technology, P.O. Box 513, 5600 MB Eindhoven, The Netherlands}
\author{Thang Ba Hoang}
\affiliation{COBRA Research Institute, Eindhoven University of Technology, P.O. Box 513, 5600 MB Eindhoven, The Netherlands}
\author{Andrea Fiore}
\affiliation{COBRA Research Institute, Eindhoven University of Technology, P.O. Box 513, 5600 MB Eindhoven, The Netherlands}


\begin{abstract}
We demonstrate a novel method to measure the $N$th-order ($N$=1, 2, 3, 4) interferometric autocorrelation with high sensitivity and temporal resolution. It is based on the combination of linear absorption and nonlinear detection in a superconducting nanodetector, providing much higher efficiency than methods based on all-optical nonlinearities. Its temporal resolution is only limited by the quasi-particle energy relaxation time, which is directly measured to be in the 20 ps range for the NbN films used in this work. We present a general model of interferometric autocorrelation with these nonlinear detectors and discuss the comparison with other approaches and possible improvements.
 
\end{abstract}



\maketitle

The temporal correlation functions of various orders are of fundamental importance in the classical and quantum description of optical fields. The first-order (field) autocorrelation function describes temporal coherence and therefore spectral linewidth, the second-order (intensity) autocorrelation is used to measure the temporal properties of pulsed sources and to distinguish quantum and classical fields, while the measurement of higher-order autocorrelation is more sensitive to coherence features (e.g. photon bunching) of the light field \cite{Aßmann2009} and can be used to determine the asymmetry of light pulses \cite{Langlois1999}. While the first-order autocorrelation function is easily measured using an interferometer and a linear detector \cite{Fox2006}, the measurement of higher-order correlation functions requires a process which is nonlinear in the intensity $I(t)$. In interferometric autocorrelators, the normalized second-order correlation function ${g}^{(2)}(\tau)\! =\! \langle I(t)I(t+\tau)\rangle/{\langle I(t) \rangle}^{2}$ is usually measured by using either second-harmonic generation (SHG) in a nonlinear crystal, followed by a linear detector \cite{Weber1967}, or two-photon absorption (TPA) in the detector itself \cite{Boitier2009,{Mollow1968}}. In both cases, the detector measures the square of the total intensity at the output of the interferometer $\langle{I}^{2}_{\textrm{tot}}(t)\rangle\! \propto\! \langle{I}^{2}(t)\rangle+2\langle I(t)I(t+{\tau}_{\textrm{d}})\rangle$ where ${\tau}_{\textrm{d}}$ is the delay between the two arms of the interferometer, together with interference terms which are sensitive to the phase properties of the beam. While these approaches offer very high temporal resolution, since the related processes are nearly instantaneous, their sensitivity is limited by the low nonlinear susceptibilities involved in the SHG or TPA process. Due to the even lower relevance of higher-order optical nonlinearities, they cannot be effectively applied to the measurement of the autocorrelations of order $N\!>\!2$. An alternative approach consists of combining linear optical detection with nonlinear processing in the electrical read-out, e.g. in a correlation card, as in the Hanbury Brown and Twiss interferometer \cite{Hanbury Brown1956}. In this case, sensitive single-photon detectors can be used, however the temporal resolution is limited $\geq100$ ps by the jitter of the detector output and of the amplification and correlation electronics \cite{StrCam}. In addition, differently from interferometric autocorrelation, these approaches do not provide any information on the phase properties.

In this work we report a novel approach to the measurement of the interferometric autocorrelation of order $N\!\geq\!2$, which is based on the combination of $linear$ $absorption$ and $nonlinear$ $detection$ in a single device. The general principle consists of absorbing incident photons in a linear absorber (i.e. a material where the absorption probability per unit time is proportional to light intensity), which produces an output pulse only if two or more photons are absorbed within a certain time interval in the fs or ps range. We show the implementation of this concept in a superconducting nanodetector, where the nonlinearity is widely tunable by varying the bias current. We directly measure the temporal dynamics of the nonlinearity in the ps range and we attribute it to the relaxation dynamics of photo-created quasi-particles (QP). We show its application to the measurement of up to fourth-order interferometric autocorrelation, observing an extremely high sensitivity related to the linear absorption process and to the low detector noise. 

\begin{figure}[!htb]
\vspace{0pt}
\centering
\begin{minipage}{0.5\textwidth}
\hspace*{0pt}
\centering
\includegraphics[width=0.5\linewidth]{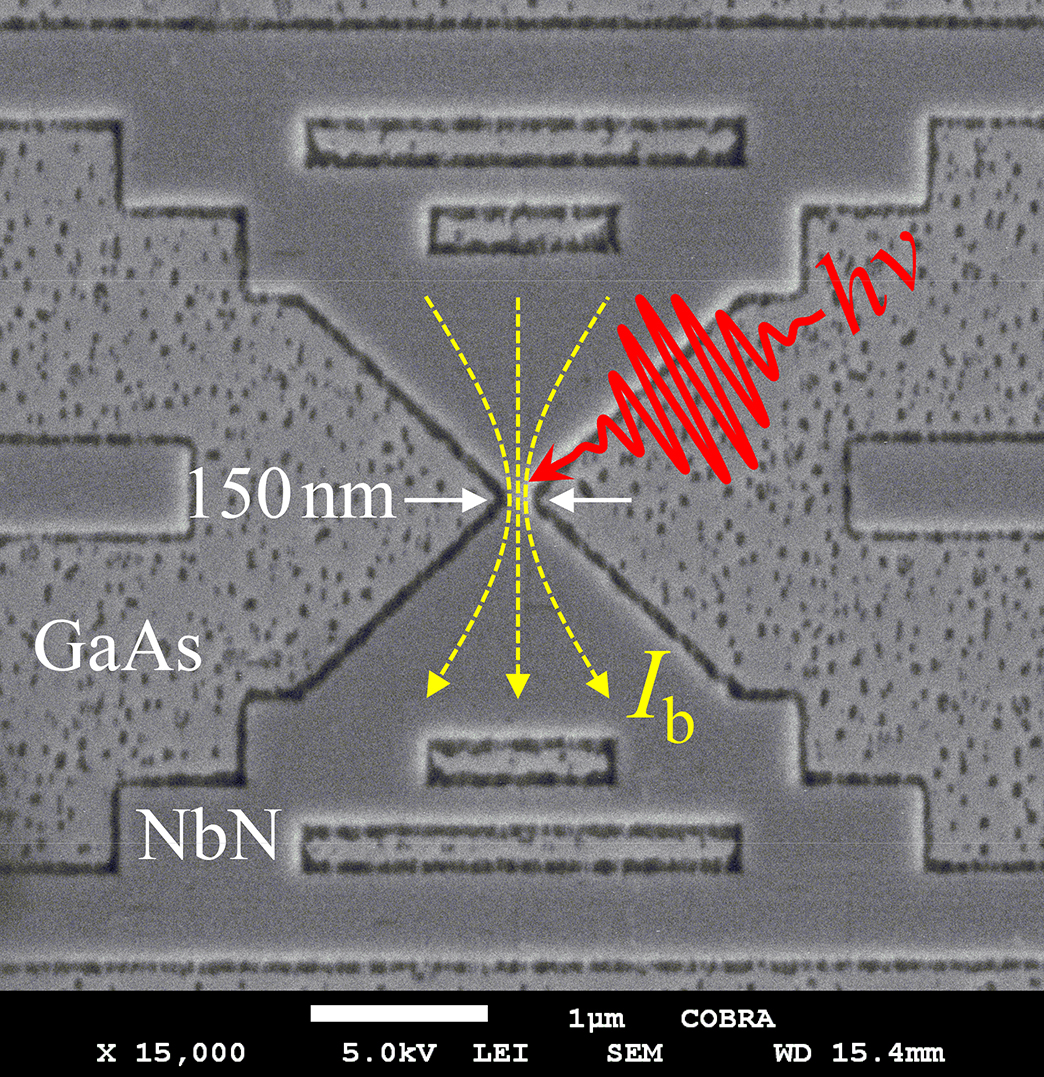}
\end{minipage}
\caption{Scanning electron microscope image of a NbN superconducting nanodetector.  
\label{fig:Graph1}}
\end{figure}

The superconducting nanodetector consists of a nanoscale constriction in a superconducting wire (see Fig.\ref{fig:Graph1}). The device used in this work is based on a 4.3 nm-thick NbN film (critical temperature ${T}_{\textrm{c}}$=10.2 K) sputtered on GaAs substrate and has a constriction size of about 150 nm, patterned by electron beam lithography and reactive-ion etching. The nanodetector is biased with a current ${I}_{\textrm{b}}$ smaller than the superconducting critical current ${I}_{\textrm{c}}$. Similarly to nanowire superconducting single-photon detectors \cite{Gol’tsman2001}, the absorption of one or more photons produces a non-equilibrium population of QP in the nanodetector's active region, locally suppressing the superconductivity and increasing the probability of vortex crossing, which can result in the transition to the normal state \cite{Semenov2008, {Bulaevskii2012}}. The detection probability is a strong function of the absorbed energy (i.e. to the number of photons) and of the bias current ${I}_{\textrm{b}}$, so that the nanodetector can be set to respond to $\geq N$ photons by choosing ${I}_{\textrm{b}}$ \cite{Bitauld2010}. A full tomographic characterization of the multi-photon response of the nanodetector under illumination with short pulses has been reported in Ref. \cite{Renema2012}. 

In this study we focus on the temporal characteristics of the nanodetector's multi-photon response and on its application as an interferometric autocorrelator. As we focus here on the application to the characterization of classical light sources, we describe the detection process as a sequence of single-photon absorption events using the semi-classical photo-detection theory. Assuming an incident light pulse with cycle-average intensity $I_{\textrm{in}}(t)$ on the nanodetector, the probability of creating a hot-spot within a time interval $\left[t+\textrm{d}t\right]$ is equal to $\zeta\cdot S\cdot {I}_{\textrm{in}}(t)\cdot \textrm{d} t$, where $\zeta$ is related to the absorptance ${\eta}_{\textrm{abs}}$ by $\zeta={\eta}_{\textrm{abs}}/h\nu$ and $S$ is the active area. In the two-photon regime, for example, the click probability ${P}_{\textrm{click}}$ (assumed $\!\ll\! 1$) is equal to the probability that two photons are absorbed, weighted by a function $\eta({\tau}_{12})$ of the time difference ${\tau}_{12}$ between absorption of the first and the second photon at times ${t}_{1}$ and ${t}_{2}$ respectively,

\vspace*{-10pt}
\begin{eqnarray}
{P}_{\textrm{click}}=\int\limits_{-\infty}^{+\infty}\int\limits_{-\infty}^{+\infty}\eta({t}_{2}-{t}_{1}) \zeta  S  {I}_{\textrm{in}}({t}_{1}) \zeta  S  {I}_{\textrm{in}}({t}_{2}) \cdot \textrm{d} {t}_{1} \cdot \textrm{d}{t}_{2}
\nonumber
\hspace*{10pt}
\\  ={\zeta}^{2}{S}^{2}\int\limits_{-\infty}^{+\infty}\eta({ \tau}_{12} ) \int\limits_{-\infty}^{+\infty} {I}_{\textrm{in}}(t) {I}_{\textrm{in}}(t+{\tau}_{12}) \cdot \textrm{d} t \cdot \textrm{d} {\tau}_{12},
\hspace*{5pt}
\label{eq:eq1}
\end{eqnarray}

The $\eta({ \tau}_{12} )$ , named nonlinear response function (NRF) in the following, depends on the QP dynamics in the superconductor and determines the intrinsic response time ${\tau}_{\textrm{ND}}$ of the nanodetector. The $\eta({ \tau}_{12} )$ is expected to decay from a maximum value of $\eta(0)$(which depends on ${I}_{\textrm{b}}$) to 0 for ${\tau}_{12}\gg {\tau}_{\textrm{ND}}$. In general, the detection probability in the $N$-photon regime ($N\geq 2$) is,

\vspace*{-15pt}
\begin{eqnarray}
{P}_{\textrm{click}}={\zeta}^{N}{S}^{N} \int\limits_{-\infty}^{+\infty}{\eta}_{N}({\tau}_{1N},{\tau}_{2N},\cdots,{\tau}_{N-1,N})\int\limits_{-\infty}^{+\infty} {I}_{\textrm{in}}(t)
\nonumber
\hspace*{20pt}
\\  {I}_{\textrm{in}}(t+{\tau}_{1N} ) {I}_{\textrm{in}}(t+{\tau}_{2N})\cdots {I}_{\textrm{in}}(t+{\tau}_{N-1,N})  
\nonumber
\hspace*{33pt}
\\   \cdot  \textrm{d} t \cdot \textrm{d} {\tau}_{1N}\cdot \textrm{d} {\tau}_{2N} \cdots \textrm{d} {\tau}_{N-1,N}               
\hspace*{96pt}  
\label{eq:eq2}                                                     
\end{eqnarray}

Where ${\tau}_{iN}={t}_{N}-{t}_{i}$ denotes the difference between the absorption times of the $i$th- and the last photon. 

When the incident pulse width is much larger than ${\tau}_{\textrm{ND}}$, Eq.(\ref{eq:eq2}) is approximated as ${P}_{\textrm{click}}\!\propto\!\int_{-\infty}^{+\infty} {I}^{N}_{\textrm{in}}(t) \cdot \textrm{d} t$. When a nanodetector is placed at the output of a Michelson interferometer, the input intensity reads $I_{\textrm{in}}(t)\!\propto\!{\left| E(t)+E(t+{\tau}_{\textrm{d}}) \right|}^{2}$ and the ${P}_{\textrm{click}}$ in $N$-photon regime is proportional to the $N$th-order interferometric autocorrelations given by ${P}_{\textrm{click}} ({\tau}_{\textrm{d}}) \propto \int_{-\infty}^{+\infty} {[{\left| E(t)+E(t+{\tau}_{\textrm{d}}) \right|}^{2}]}^{N} \!\!\!\cdot \textrm{d} t $. In the two-photon regime, by filtering out the interference terms we obtain ${P}_{\textrm{click}}({\tau}_{\textrm{d}}) \!\propto\! \langle{I}^{2}(t)\rangle+2\langle I(t)I(t+{\tau}_{\textrm{d}})\rangle$, and therefore the ${g}^{(2)}({\tau}_{\textrm{d}})$, similarly to the SHG- and TPA-based autocorrelators.

The NRF can be measured by probing the autocorrelator with short pulses. For pulse duration much shorter than ${\tau}_{\textrm{ND}}$, the response of the nanodetector placed at the output of a Michelson interferometer can be found from Eq.(\ref{eq:eq2}). In the two-photon regime for example, after filtering out the interference terms, Eq.(\ref{eq:eq1}) becomes ${P}_{\textrm{click}}({\tau}_{\textrm{d}})\!\propto\! \eta(0)+\eta({\tau}_{\textrm{d}})+f( {\tau}_{\textrm{d}})$ \cite{SM}. The function $f( {\tau}_{\textrm{d}})$ depends on the degree of first-order coherence of the input light and is different from zero only for delays shorter than the coherence time. For longer delays, the ${P}_{\textrm{click}}({\tau}_{\textrm{d}})$, normalized by its value at ${\tau}_{\textrm{d}}\gg {\tau}_{\textrm{ND}}$, is expected to vary as ${P}_{\textrm{click}}({\tau}_{\textrm{d}})/{P}_{\textrm{click}}(\infty)\!=\!1+\eta( {\tau}_{\textrm{d}})/\eta(0)$, so that the NRF and therefore the  autocorrelator's timing resolution can be measured.

\begin{figure}[!htb]
\vspace{0pt}
\centering
\begin{minipage}{0.5\textwidth}
\hspace{-15pt}
\centering
\includegraphics[width=0.7\linewidth]{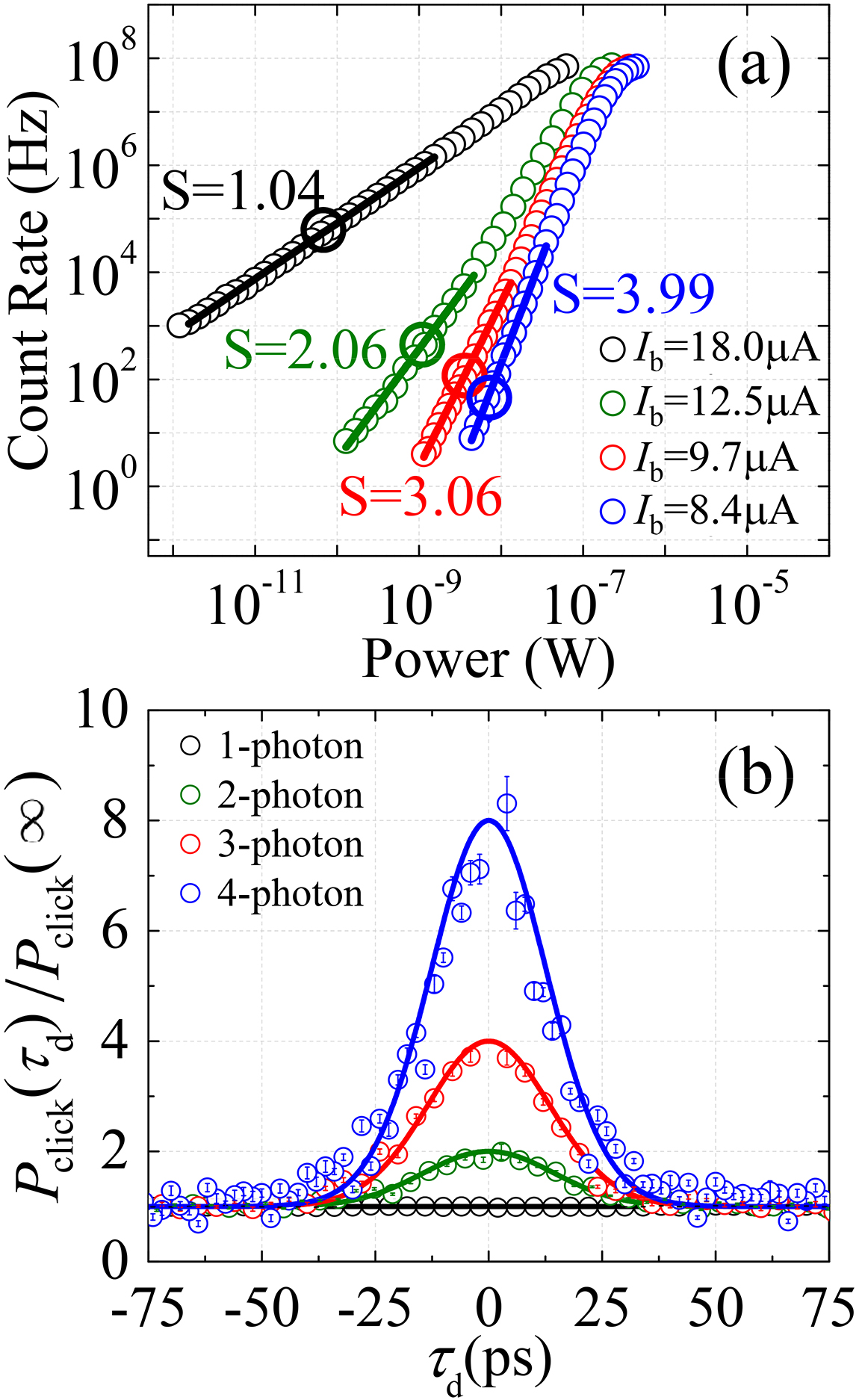}
\end{minipage}
\caption{(a) CR as a function of the light power for different ${I}_{\textrm{b}}$. The fitting lines with slopes of 1.04, 2.06, 3.06 and 3.99 indicate the one-, two-, three- and four-photon regimes respectively. (b) Normalized CR (circles) as a function of ${\tau}_{\textrm{d}}$, measured when the detector was working at the four chosen points (marked by large circles in (a)). The solid lines are calculation results (see text). 
\label{fig:Graph2}}
\end{figure}

The NRF was first measured by sending 1.6 ps pulses from an optical parametric oscillator (OPO) at $\lambda\!=\!1.13\mu m$ into a fiber-based Michelson interferometer and then to a nanodetector held at a temperature of 1.2 K using a lensed fiber producing a spot with an ${e}^{-2}$ diameter of  5 $\mu$m. The delay ${\tau}_{\textrm{d}}$ in the interferometer is controlled by a motorized delay line (coarse control) and a fiber stretcher (fine control).  The ${I}_{\textrm{c}}$ of the device was about 26 $\mu$A. During the measurement, the nanodetector was set in different photon regimes by choosing a proper ${I}_{\textrm{b}}$. As shown in Fig. \ref{fig:Graph2}(a), the count rate (CR) was measured as a function of the light power at different ${I}_{\textrm{b}}$. The solid lines with slopes of 1.04, 2.06, 3.06 and 3.99 are the fits to the measured data in log-log scale, in the power ranges where the one-, two-, three- and four-photon response was dominant \cite{Bitauld2010}. We chose one point at each of the four photon regimes, and measured the CR as a function of ${\tau}_{\textrm{d}}$, normalized by its values at long delays, as shown in Fig. \ref{fig:Graph2}(b). The data points near ${\tau}_{\textrm{d}}\!=\!0$, where the measured autocorrelation is sensitive to the first-order coherence, are not shown in the plot and were not considered in the fit since they introduce additional fitting error \cite{SM}. At ${I}_{\textrm{b}}\!=\!18.0$ $\mu$A, the CR is independent of the delay, apart from the short-delay interference fringes, as expected, since the detector is working in the linear regime. When the ${I}_{\textrm{b}}$ was lowered to 12.5 $\mu$A, 9.7 $\mu$A and 8.4 $\mu$A (corresponding to the two-, three- and four-photon regime, respectively), a maximum is observed at zero delay. Since the width of these peaks is much wider than the OPO pulse width, the measurement probes the intrinsic response of the detector. In particular, the two-photon normalized response in Fig. \ref{fig:Graph2}(b) directly provides the $1+\eta({\tau}_{\textrm{d}})/\eta(0)$ dependence. The $\eta(\tau)$ is determined by the thermalization and relaxation processes of the photo-created QP and by the functional dependence of the  ${P}_{\textrm{click}}$ on the QP concentration \cite{Semenov2008, {Bulaevskii2012}}. Indeed, the QP population first grows as the electron population thermalizes via electron-electron scattering, in a timescale of few ps, then decays due to electron-photon interaction and phonon escaping to the substrate \cite{Semenov2001}. This QP decay is expected to determine the ${\tau}_{\textrm{ND}}$ in the $N$-photon regime, since the ${I}_{\textrm{b}}$ is chosen so that  ${P}_{\textrm{click}}$ is high only when all QP produced by the $N$ photons are present at the same time. As ${P}_{\textrm{click}}$ is a strong function of the QP number, involving many microscopic parameters, a fit of the measured $\eta(\tau)$ using a microscopic model would not be reliable. Instead, we introduce an empirical Gaussian NRF defined as $\eta({ \tau}_{12})\!=\!\eta(0)\textrm{exp}[-{({ \tau}_{12}/{\tau}_{\textrm{ND}})}^{2}]$. The fits to ten measured two-photon autocorrelation traces provide a ${\tau}_{\textrm{ND}}$ value of $20.4\pm0.8$ ps. In order to fit the $N\!>\!2$ traces we further assume that the multi-photon response factorizes as ${\eta}_{N}({\tau}_{1N},{\tau}_{2N},\cdots,{\tau}_{N-1,N})\!=\!\eta({\tau}_{1N})\eta({\tau}_{2N})\cdots \eta({\tau}_{N-1,N})$. This is reasonable if one assumes that ${\eta}_{N}$ has an approximately exponential dependence on the total QP concentration after absorption of the $N$th-photon, as suggested by the vortex-assisted photo-detection model \cite{Semenov2008,Bulaevskii2012}, and that the QP relaxation time does not depend on QP concentration. Using the ${\tau}_{\textrm{ND}}$ value extracted from the two-photon autocorrelation traces as described above, ${P}_{\textrm{click}}({\tau}_{\textrm{d}})$ was calculated for the three- and four-photon regimes from Eq.(\ref{eq:eq2}) without additional fitting parameters and shows excellent agreement with the experiment (Fig. \ref{fig:Graph2}(b)), which provides a strong experimental support to our model. The value of ${\tau}_{\textrm{ND}}$ measured here at 1.2K is comparable but shorter than the hot-electron energy relaxation time in NbN microbridge measured by electro-optic sampling \cite{Il’in2000} and by Terahertz spectroscopy \cite{Beck2011} - this difference may be related to the difference in film thickness and film properties. 

\begin{figure*}
\vspace{0pt}
\centering
\hspace*{-5pt}
\centering
\includegraphics[height=0.21\textwidth]{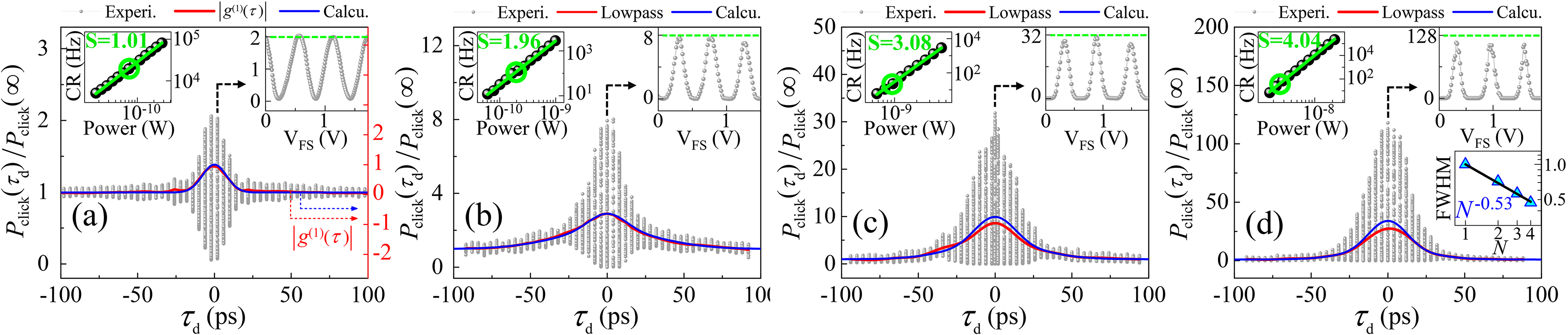}
\caption{(a-d) Normalized CR as a function of ${\tau}_{\textrm{d}}$ in the different regimes at ${I}_{\textrm{b}}$ of 18.0 $\mu$A, 14.0$\mu$A,11.5$\mu$A and 9.6$\mu$A respectively. The $|{g}^{(1)}({\tau}_{\textrm{d}})|$ (red line in (a)) and low-pass curves (red lines in (b-d)) agree well with the calculations (blue lines). Upper-left insets: CR as a function of incident power in log-log scale. The fitting slopes of 1.01, 1.96, 3.08 and 4.04 indicate the one-, two-, three- and four-photon regimes, respectively. The green circles indicate the points chosen for the measurements in the main panel. Upper-right insets: Expanded view of the interference fringes at small delays, plotted as a function of the voltage ${V}_{\textrm{FS}}$ applied to the fiber stretcher.  Lower-right inset in (d): FWHM of the fringes normalized by their period extracted from the upper-right insets in (a-d) as a function of $N$, showing a $1/\sqrt{N}$ dependence.  
\label{fig:Graph3}}
\end{figure*}

With the knowledge of the temporal resolution, autocorrelation experiments have been performed on pulses generated by a gain-switched 1.3$\mu$m diode laser with 10MHz repetition rate and about 70ps pulse width. The one-, two-, three- and four-photon regimes were first found by choosing different ${I}_{\textrm{b}}$. At each ${I}_{\textrm{b}}$, CR were recorded as a function of ${\tau}_{\textrm{d}}$ in the $N$-photon ($N$=1, 2, 3, 4) regime as shown in Fig. \ref{fig:Graph3}(a-d), respectively. The fringe contrast ratio is observed to increase with $N$, in good agreement with the theoretical values of 2, 8, 32 and 128 for $N$\!=\!1, 2, 3 and 4, respectively. In the one-photon regime (Fig. \ref{fig:Graph3}(a)), the normalized first-order autocorrelation function ${g}^{(1)}({\tau}_{\textrm{d}})$ was calculated from the visibility of the interference fringes \cite{Fox2006}. For $N\!>\!1$, higher-order intensity autocorrelation traces were obtained by applying a low-pass filter to the interferograms in Fig. \ref{fig:Graph3}(b-d). In order to explain the experimental results, we assumed that the incident light was a Gaussian pulse with a linear chirp. The electric field is modeled as $E(t)\!=\!{E}_{0}\textrm{exp}[-4\textrm{ln}(2)(1+iA){t}^{2}/{\tau}^{2}_{\textrm{p}}]$, where ${E}_{0}$ is the field amplitude, $A$ is the linear chirp parameter \cite{Diels1985} and $\tau_{\textrm{p}}$ is the full width at half maximum (FWHM) of the pulse. By fitting the measured $|{g}^{(1)}({\tau}_{\textrm{d}})|$ in Fig. \ref{fig:Graph3}(a) together with the second-order intensity autocorrelation (low-pass trace) in Fig. \ref{fig:Graph3}(b), $A$ and $\tau_{\textrm{p}}$ were determined to be 5.3 and 70.6 ps. The third and fourth-order interferometric autocorrelations were then calculated based on Eq.(\ref{eq:eq2}) without additional fitting parameters. Their low-pass traces show a good agreement with the experiments, considering the very simplified assumption for the chirp. An enlarged view of each interferogram is shown as upper-right inset in each panel of Fig. \ref{fig:Graph3}, showing a clear narrowing of the fringes for increasing $N$. As shown in the lower-right inset of Fig. \ref{fig:Graph3}(d), the FWHM of the fringes, normalized by their period, scales as approximately $1/\sqrt{N}$, which is a characteristic of multi-photon interferometry \cite{Yablonovitch1999} and further confirms our conclusions.

A comparison of the sensitivity of our nanodetector to the detection schemes used in conventional autocorrelators provides useful insight. For input pulses longer than $\tau_{\textrm{ND}}$, Eq.(\ref{eq:eq2}) can be written as ${P}_{\textrm{click}}\!=\!{C}_{\textrm{ND}} \int_{-\infty}^{+\infty} {P}^{2}_{\textrm{in}}(t) \textrm{d} t$, where ${P}_{\textrm{in}}\!\!=\!\!{I}_{\textrm{in}}S$ is the incident power and ${C}_{\textrm{ND}}\!=\!\sqrt{\pi}{\eta }^{2}_{\textrm{abs}}\eta(0){\tau}_{\textrm{ND}}/{(h\nu)}^{2}$ represents a nonlinear response efficiency. A similar expression for ${P}_{\textrm{click}}$ can be derived for nonlinear detectors based on SHG or TPA, where the corresponding ${C}_{\textrm{SHG}}$ and ${C}_{\textrm{TPA}}$ values are related to the SHG nonlinear conversion efficiency and to the TPA coefficient ${\beta}_{2}$, respectively. In all three cases, the sensitivity can be defined by imposing that the CR is equal to the dark count rate ${R}_{\textrm{dark}}$ in the detector. For a periodic sequence of square pulses with repetition rate ${f}_{\textrm{rep}}$ the condition ${f}_{\textrm{rep}}{P}_{\textrm{click}}\geq {R}_{\textrm{dark}}$ leads to  ${P}_{\textrm{pk}}{P}_{\textrm{av}}\!\geq\! {R}_{\textrm{dark}}/C$, where ${P}_{\textrm{pk}}$ and ${P}_{\textrm{av}}$ are the peak and average power, respectively. The two-photon autocorrelation trace in Fig. \ref{fig:Graph3}(b) was taken at  ${P}_{\textrm{pk}}{P}_{\textrm{av}}\!=\!5.6\times {10}^{-17} {\textrm{W}}^{2}$, about seven orders of magnitude lower than the minimum reported  ${P}_{\textrm{pk}}{P}_{\textrm{av}}$ for TPA using a GaAs PM tube \cite{Roth2002} and about two orders of magnitude lower than the lowest  ${P}_{\textrm{pk}}{P}_{\textrm{av}}$ reported in the second-order interferometric correlations using SHG \cite{Hsu2011}. Using the measured values of ${\eta}_{\textrm{abs}}\!=\!1.5\times {10}^{-4}$,  $\eta(0)\!\approx\!0.5$ \cite{Renema2012} and ${R}_{\textrm{dark}}\!=\!1\textrm{Hz}$, we derive a sensitivity of ${P}^{\textrm{min}}_{\textrm{pk}}{P}^{\textrm{min}}_{\textrm{av}}\!\approx\! 5.8\times {10}^{-20} {\textrm{W}}^{2}$, corresponding to $\sim$4 photons per pulse in our experiment.  In higher-photon regime the advantage of using linear absorption is even larger. Indeed, the three-photon autocorrelation shown in Fig. \ref{fig:Graph3}(c) was performed at an average power of about 1nW, corresponding to ${P}_{\textrm{pk}}^{2}{P}_{\textrm{av}}\!\approx\! 2.0\times {10}^{-21} {\textrm{W}}^{3}$, an improvement of about twenty-one orders of magnitude over Ref. \cite{Wei2011}. $N$-photon interferometric autocorrelation for $N\!>\!3$ has not been reported before to the best of our knowledge.

The very high nonlinear response of the nanodetector, can be directly traced to the finite size and time duration of the hot-spot created by the $real$ absorption of one photon, as compared to virtual transitions involved in TPA. Indeed, a two-photon detection is triggered if the second photon is absorbed within the volume and time duration of the hot-spot created by the first photon. This shows that a compromise exists between ${C}_{\textrm{ND}}$ and ${\tau}_{\textrm{ND}}$: for the nanodetector the ${\tau}_{\textrm{ND}}$ is determined by the QP relaxation time, while in TPA it is related to the lifetime of the virtual states associated to the TPA transition, of the order of fs \cite{Boitier2009}. A similar compromise exists in SHG-based autocorrelators, where higher conversion efficiency requires a longer SHG crystal translating into a smaller phase-matching bandwidth and lower temporal resolution \cite{Weiner1983, Yang2007}. Interestingly, in an absorption thickness of 4.3 nm the nanodetector reaches a nonlinear efficiency $C\!\approx\! 1.7\times {10}^{19} {\textrm{W}}^{-1}{\textrm{J}}^{-1}$, several orders of magnitude higher than $\mu$m-thick TPA absorbers and comparable to cm-long SHG crystals, showing the giant effective nonlinearity achieved through the combination of linear absorption and nonlinear detection. We note that a key advantage as compared to SHG-based autocorrelators is the wide optical bandwidth, limited only by the requirement to operate in the desired $N$-photon regime, which can be easily adjusted by varying the ${I}_{\textrm{b}}$. This in principle enables the measurement of interferometric autocorrelation from the visible to the mid-infrared.

The sensitivity of our autocorrelator is presently limited by the low ${\eta}_{\textrm{abs}}$, related to the spatial mismatch between the incoming beam and the nanodetector's active area and to the small thickness of the NbN film. By focusing the beam with a high-numerical aperture lens, a much higher absorptance ${\eta}_{\textrm{abs}}\!\approx\!{10}^{-2}$ can be achieved \cite{lens}. The integration of a plasmonic antenna and of a bottom reflector \cite{Gaggero2010} could increase ${\eta}_{\textrm{abs}}$ to the ${10}^{-1}$ range, leading to ${P}^{\textrm{min}}_{\textrm{pk}}{P}^{\textrm{min}}_{\textrm{av}}\!\approx\! {10}^{-25} {\textrm{W}}^{2}$ for the two-photon autocorrelation. On the other hand, increasing the detector area (as done in meander nanowire detectors \cite{Verevkin2002}) results in ${\eta}_{\textrm{abs}}\!\propto\! L$ and $\eta(0)\!\propto\! {l}_{\textrm{hs}}/L$ ($L$ is the nanowire length, ${l}_{\textrm{hs}}$ is the hot-spot length) so that ${C}_{\textrm{ND}}$ scales linearly with ${\eta}_{\textrm{abs}}$. The nanoscale nature of the detector is therefore crucial to reaching the ultimate sensitivity. Finally, we note that the 20ps temporal resolution in our experiments is limited to the QP relaxation time in NbN films and could be much improved using different superconducting materials, such as high-${T}_{\textrm{c}}$ Y-Ba-Cu-O films, where relaxation times $\sim$1 ps were observed \cite{Lindgren1996}, opening the way to the characterization of $N$th-order correlation functions in few-ps range with unprecedented sensitivity.

\begin{acknowledgements}
The authors gratefully thank Prof. E. Rosencher and J. J. Renema for enlightening discussions, D. Sahin for taking the SEM picture (Fig. \ref{fig:Graph1}), J. E. M. Haverkort and S. Mokhlespour for assistance in operating the OPO system, and T. Xia and F.M. Pagliano for technical help. This work is part of the research programme of the Foundation for Fundamental Research on Matter (FOM), which is financially supported by the Netherlands Organisation for Scientific Research (NWO), and is also supported by NanoNextNL, a micro and nanotechnology program of the Dutch ministry of economic affairs and agriculture and innovation (EL$\&$I) and 130 partners, and by the European Commission through FP7 project Q-ESSENCE (Contract No.248095).
\end{acknowledgements}


\begin{thebibliography}{35}
\bibitem{lens}
\bibnamefont{This value is obtained by assuming the diameter of the Gaussian beam spot decreases from 5$\mu$m to 0.5$\mu$m by using high-numerical aperture lens, and the nanodetector's active area is 150nm$\times$150nm.}
\end{thebibliography}

\newpage

\noindent \textbf{\Large{Supplemental Material:}}
\newline

In the general case, the response of the nanodetector, which is set in the two-photon regime and placed at the output of the interferometer, is found using Eq. (1) and reads,

\begin{widetext}
\vspace*{-10pt}
\begin{eqnarray}
{P}_{\textrm{click}}({\tau}_{\textrm{d}})\propto 2\!\!\int\limits_{-\infty}^{+\infty}\!\!\eta({ \tau}_{12} ){G}^{(2)}({\tau}_{12} )\cdot\text{d}{\tau}_{12}+\!\!\int\limits_{-\infty}^{+\infty}\!\!\eta({ \tau}_{12} ){G}^{(2)}({\tau}_{12}+{\tau}_{\textrm{d}})\cdot\text{d}{\tau}_{12}+\!\!\int\limits_{-\infty}^{+\infty}\!\!\eta({ \tau}_{12} ){G}^{(2)}({\tau}_{12}-{\tau}_{\textrm{d}})\cdot\text{d}{\tau}_{12}
\nonumber
\hspace*{0pt}
\\ +2\textrm{Re}\left[\int\limits_{-\infty}^{+\infty}\!\!\eta({\tau}_{12})\!\!\int\limits_{-\infty}^{+\infty}\!\!{E}^{\ast}(t)E(t+{\tau}_{12})E(t+{\tau}_{\textrm{d}}){E}^{\ast}(t+{\tau}_{12}+{\tau}_{\textrm{d}})\cdot\text{d}t\cdot\text{d}{\tau}_{12}\right]
\nonumber
\hspace*{72pt}
\\ +\!\!\int\limits_{-\infty}^{+\infty}\!\!\eta({\tau}_{12} ){F}_{1}({\tau}_{12},\omega{\tau}_{\textrm{d}})\cdot\text{d}{\tau}_{12}+\!\!\int\limits_{-\infty}^{+\infty}\!\!\eta({\tau}_{12} ){F}_{2}({\tau}_{12},2\omega{\tau}_{\textrm{d}})\cdot\text{d}{\tau}_{12}  \qquad\qquad\qquad \qquad  \qquad  \qquad \qquad (\textrm{S1})
\nonumber
\hspace*{-26pt}
\end{eqnarray}      
\end{widetext}

The first term on the right side of Eq. (S1) is a constant with respect to ${\tau}_{\textrm{d}}$. The second and the third terms represent the convolution of the NRF with the second-order correlation function defined as  ${G}^{(2)}(\tau)=\!\!\int_{-\infty}^{+\infty}\!\!I(t)I(t+\tau)\cdot\text{d}t$, while the fourth term is sensitive to the phase-coherence of the input beam and gives rise to an additional peak in the filtered autocorrelation signal at small delays. The last two terms on the right side of Eq. (S1) are interference terms which vary with $\omega{\tau}_{\textrm{d}}$ or $2\omega{\tau}_{\textrm{d}}$. Since they average to zero they can be easily filtered out using a low-pass filter. 

In the case where the pulse is much shorter than ${\tau}_{\textrm{ND}}$, Eq. (S1) is written as,
\begin{widetext}
\vspace*{-10pt}
\begin{eqnarray}
{P}_{\textrm{click}}({\tau}_{\textrm{d}})\propto \eta(0)+\eta({\tau}_{\textrm{d}})+\frac{\eta(0){\left|{G}^{(1)}({\tau}_{\textrm{d}})\right|}^{2}}{\!\!\int\limits_{-\infty}^{+\infty}\!\!{G}^{(2)}({\tau}_{12} )\cdot\text{d}{\tau}_{12}}+\textrm{interference terms}\qquad \qquad (\textrm{S2})
\nonumber
\hspace*{0pt}
\end{eqnarray}      
\end{widetext}

Where ${G}^{(1)}({\tau}_{\textrm{d}})$  is the first-order correlation function defined as  ${G}^{(1)}({\tau}_{\textrm{d}})=\!\!\int_{-\infty}^{+\infty}\!\!{E}^{\ast}(t)E(t+{\tau}_{\textrm{d}})\cdot\text{d}t$. The third term on the right side of Eq. (S2), which comes from the fourth term in Eq. (S1), depends on the first-order coherence properties of the input beam, but, differently from the interference terms, does not disappear upon filtering and produces a peak in the filtered autocorrelation at zero delay.

\begin{figure*}
\vspace{0pt}
\centering
\hspace*{-5pt}
\centering
\includegraphics[height=0.3\textwidth]{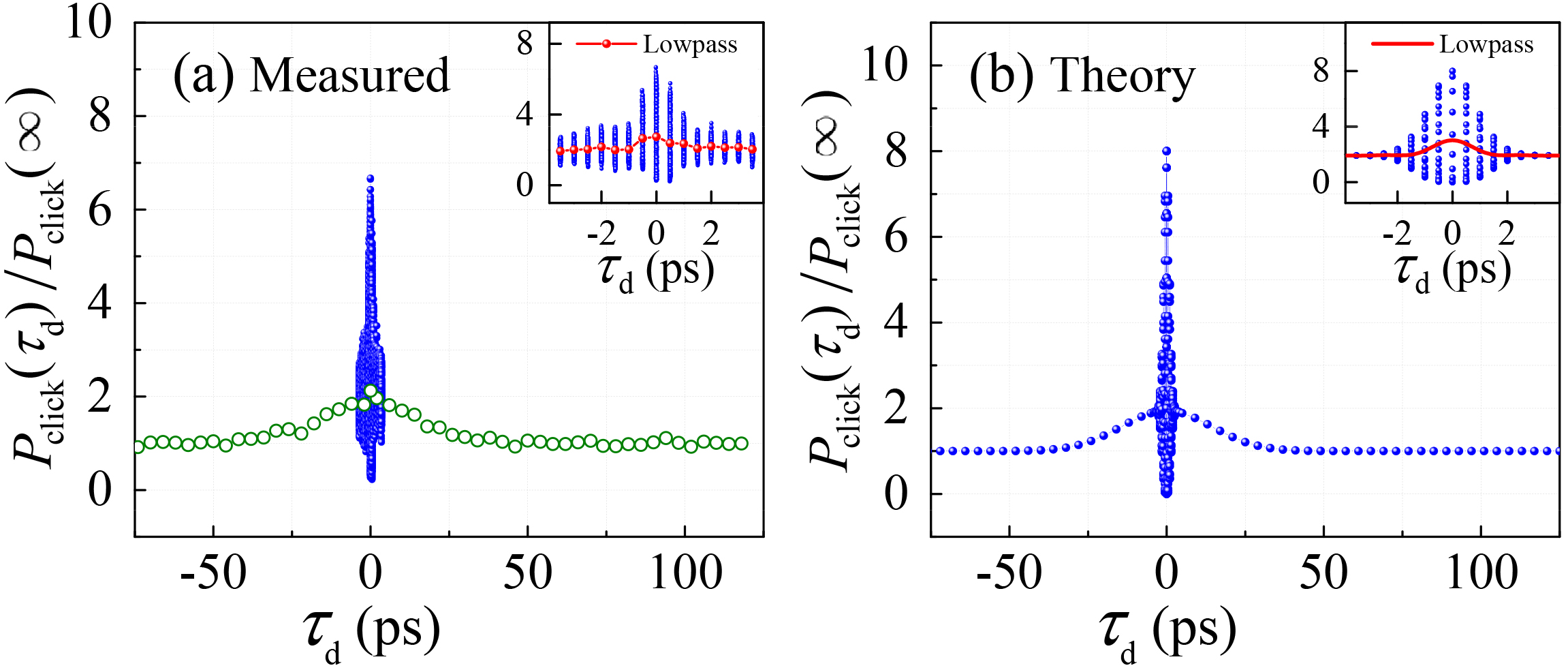}
\caption{ (a) The measured count rate as a function of the time delay, normalized by its value at long delays. (b) Simulation based on Eq. (S1). An enlarged view at small delays is shown in the insets.  
\label{fig:Fig4}}
\nonumber
\end{figure*}
\nonumber

In order to confirm this prediction, we studied the short delay window in the experiment where the OPO short pulses (1.6ps, 1.13$\mu$m) were used to characterize the NRF of the nanodetector in the two photon regime. To this aim, the delay ${\tau}_{\textrm{d}}$ was controlled by a motorized delay line (coarse control) and a fiber stretcher (FS) (fine control). A driving voltage ${V}_{\textrm{FS}}$ (zigzag wave, $\pm$5V, 20mHz) was applied on the fiber stretcher to change the optical path difference between the two arms. At each coarse delay, the counter recorded the count rate as a function of the ${V}_{\textrm{FS}}$. Due to the short pulse width of OPO, the interference fringes were observed at small coarse delays and an additional peak appeared in a small window near zero delay (blue points as shown in Fig. 4(a), also shown in an enlarged view in the inset), on top of the broad peak related to the detector NRF, as observed in Fig. 2(b) of the main text and replotted as green circles in Fig. 4(a). A theoretical comparison based on Eq. (S1) is shown in Fig. 4(b). It considers a 20.4ps intrinsic response time of the nanodetector, and a probe pulse of 1.6ps duration and 1.13$\mu$m wavelength as used in the experiments, showing a good agreement with the measurements. We note that this narrow peak appears also in the filtered autocorrelation, shown as red lines in the insets of Fig. 4(a) and (b). However, as indicated by Eq. (S2), its width relates to the first-order correlation function, and therefore to the coherence time, and not to the pulse width. Indeed, we checked numerically that in a chirped pulse the narrow peak width scales with the chirp parameter. This confirms that ${\tau}_{\textrm{ND}}$ represents the limit to the temporal resolution for the measurement of the second-order correlation function.

\end{document}